# Masses and decay constants of the B-system in Lattice QCD [1]

C. Alexandrou

Department of Natural Sciences, University of Cyprus, Nicosia, Cyprus[2]

## Abstract

Heavy-light bound states are studied in Lattice QCD with emphasis on parameters of the B-system relevant to experiment. Results are obtained on lattices with lattice spacings from about 0.15 fm to 0.06 fm corresponding to $\beta = 5.74, 6.0$ and $6.26$, and covering sizes from about 0.7 fm to 2 fm. From our results at the infinite quark mass limit and from propagating heavy quarks with mass of about 1-2.5 GeV we extrapolate to the b-quark mass to obtain the decay constant of the B-meson as well as the mass of $\Lambda_b$ in the continuum. The necessary extrapolations introduce rather large errors and lead to the value $f_B = 180(50)$ MeV. We compare with the results from other lattice groups as well as with predictions coming from sum rules. The phenomenological consequences of this value of $f_B$, together with new experimental constraints, are briefly discussed.

## 1. Introduction

The extraction of Cabibbo-Kobayashi-Maskawa (CKM) matrix elements from experimental measurements is hampered by our poor knowledge of hadronic matrix elements. Therefore, in addition to increasing the experimental accuracy of the results, we also need non-perturbative techniques to evaluate the hadronic contributions. In particular, weak decays of heavy hadrons play a key role in probing the hadronic sector of the Standard Model [1] and the big challenge is to establish direct CP-violation effects in B decays, as they are predicted in the Cabibbo-Kobayashi-Maskawa mechanism [2, 3]. Within this scheme, by virtue of the unitarity of the CKM matrix, the CP-violation effects, in principle, can be predicted entirely in terms of CP-conserving experimental quantities. At present the most important constraints for the CP-violating CKM phase $\delta$ come from the observed CP-violation in the $K^0 - \bar{K}^0$ system and from results on $B^0 - \bar{B}^0$ mixing and semileptonic B-decays [4, 5]. For the prediction of $\delta$, one needs the values of the leptonic B-meson decay constant $f_B$ and the $B_B$ factor, which are not known experimentally. Theoretical predictions for heavy-light bound states like the B-meson, have been notoriously difficult and have in the past produced a rather big range of values. They were based on QCD sum rules [6], various potential models [7] and lattice calculations at the D-meson mass which were then extrapolated to the B- meson, assuming the scaling law [8]

$$f_P(M_P) \sqrt{M_P} = \text{constant}$$

---

[1] Invited talk given at the Bogoliubov International Symposium, Dubna, Russia, 18-23 August 1994.

[2] Partly supported by the Paul Scherrer Institute, CH-5232 Villigen, Switzerland.

to hold. Usually a range 100 MeV $\leq f_B \leq$ 200 MeV had been anticipated in phenomenological discussions [3, 5].

Needless to say, a full lattice determination of $f_B$, being an *ab initio* calculation from the QCD Lagrangian, would be superior to any other way of predicting this quantity. Practical lattice simulations of QCD are faced, however, with a number of problems that must be carefully dealt with before one can claim reliable results from this theoretical approach. For B-mesons, due to their large mass, the limited lattice resolution (*finite lattice constant a*) is potentially the most dangerous source of a systematic error: with quarks treated as Wilson fermions, as we do throughout this work, such scaling violations are first-order effects in $a$. In the instance of *heavy-light* quark systems on the lattice, $\mathcal{O}(a)$ terms translate into $\mathcal{O}(a M_i)$ effects and with present lattice resolution limited to lattice spacings $a \geq (4 GeV)^{-1}$, we have to cope with the problem that $a M_B > 1$.

In the past couple of years we have learned how to deal in a systematic way with heavy-light systems. The infinite heavy quark mass limit is investigated in depth by several groups [9, 10, 11, 12]. Smearing techniques have proved essential in extracting reliably the ground state with the Fermilab group [13] having developed a method to construct the most appropriate mesonic wavefunctions. Allton recently analysed [14] the available static results for the decay constant $f_B^{\text{stat}}$ and found that scaling sets in at $\beta \gtrsim 6.0$. At the same time a number of calculations were performed with heavy propagating quarks [11, 12, 15, 16] approaching the $b-$ quark from the lower mass side. Investigation of the dependence of $f_P$ on the heavy quark mass is necessary in order to extrapolate to the B-meson mass. This mass extrapolation, together with the continuum extrapolation, introduces the major source of uncertainty on our lattice results. Use of an improved action with only $\mathcal{O}(a^2)$ effects [17] ablates finite $a$ contamination and it is in this direction that future progress is to be expected.

In this talk we discuss the techniques we used to obtain a lattice estimate for $f_B$ as well as for the mass of $\Lambda_b$. Since this talk is addressed also to non-lattice specialists we start with a brief introduction of lattice QCD with emphasis on the relevant features for a numerical calculation. We will review the static approximation [18] and compare results from various groups as well as with results from QCD sum rules. This will be followed with a discussion of the results using propagating heavy quarks and of our procedure to obtain the continuum limit. We will also comment on the calculation of the $\Lambda_b$ mass [19]. In the last section we will see that the lattice errors on $f_B$, $B_K$ and $B_B$ are still large and these, in combination with the experimental uncertainties, make predictions of the unitarity triangle rather limited.

## 2. Introduction to Lattice QCD

The QCD Langragian density for pure gauge is given by

$$\mathcal{L}_U = -\frac{1}{4} F_{\mu\nu}^c F^{\mu\nu c} \tag{2.1}$$

where the greek (roman) letters denote Dirac (colour=1,2,...,8) indices and

$$F_{\mu\nu}^c = \partial_\mu A_\nu^c - \partial_\nu A_\mu^c + g_0 f^{cde} A_\mu^d A_\nu^e$$

with $f^{cde}$ the SU(3) structure constants.

In lattice QCD we consider a discretized space-time with lattice spacing $a$ as shown in fig. 1.

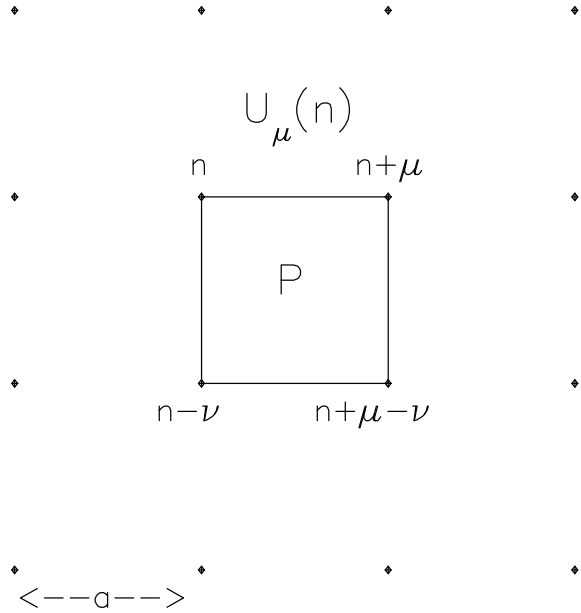

Figure 1: Link variables $U_\mu(n)$ defined on a lattice of spacing $a$

To put pure gauge QCD on the lattice preserving gauge invariance, the appropriate variables to use are the link variables $U_\mu(n)$ [20]

$$U_\mu(n) = \exp(iag_0 A_\mu(n)) \qquad (2.2)$$

which connect side $n$ along the $\mu$ - direction with the neighbouring side $n + \hat{\mu}$. A plaquette $P$ is defined as the product $U_P$ of four links round an elementary square. All space is thus covered with such plaquettes and the lattice action is the sum of them all.

$$S_U = \beta \sum_P \left[ 1 - \frac{1}{6} \text{Tr} \left( U_P + U_P^\dagger \right) \right] \qquad (2.3)$$

where

$$U_P = U_\mu(n) U_{-\nu}(n + \hat{\mu}) U_{-\mu}(n + \hat{\mu} - \hat{\nu}) U_\nu(n - \hat{\nu}) \qquad (2.4)$$

That eq.(2.4) reduces to (2.1) can be easily demostrated, by taking the naive continuum limit $a \to 0$. We have introduced $\beta = 6/g_0^2$ which is an input in a lattice calculation and it is the only parameter for pure gauge. It determines the coupling constant and therefore the lattice spacing $a$. The continuum limit is recovered by letting $g_0 \to 0$. Present lattice calculations are done in the range $6.0 \lesssim \beta \lesssim 6.5$ which translate into lattice spacing of about (2 GeV )$^{-1}$ to (4 GeV) $^{-1}$.

Putting the quarks on the lattice is a much more complicated task that the gauge fields. We will not dwell on this problem here but just remark that there is no completely satisfactory way of treating fermions, with each known method providing a partial solution to the problem. We use the so called Wilson fermions where the doubling problem is avoided at the expence of explicit breaking of chiral symmerty [21]. The action with the quarks is then taken to be

$$S_F^W = \sum_{x,y} \bar{q}(x) M_W(x,y) q(y) \qquad (2.5)$$

$$= \sum_x \left\{ \bar{q}(x)q(x) - \kappa \sum_\mu \left( \bar{q}(x)(1-\gamma_\mu) U_\mu(x) q(x+\hat{\mu}) + \bar{q}(x+\hat{\mu})(1+\gamma_\mu) U_\mu^\dagger(x) q(x) \right) \right\}$$

with $M_W(x,y)$ the fermionic matrix for Wilson fermions. The hoping parameter $\kappa$ determines the naive quark mass via

$$m_q = \frac{1}{2a} \left( \frac{1}{\kappa} - \frac{1}{\kappa_c} \right) \quad . \tag{2.6}$$

It is known from chiral perturbation theory [22] that $(am_\pi)^2 \propto am_q$ and therefore the chiral limit is recovered at the critical $\kappa$ value, $\kappa_c$, which is determined dynamically in the lattice calculation. In other words breaking of chiral symmetry introduces a fine tuning to be performed before lattice results are to match the continuum theory.

Up to now the theory is defined on an infinite lattice. In numerical calculations one has to work on a *finite* lattice of spatial length $L = aN_S$ and temporal length $T = aN_T$ with $N_S$, $N_T$ the number of spatial and temporal sites respectively. Therefore in addition to finite-$a$ effects one has to consider finite volume contamination. Usually this is checked by performing the calculation at the same $\beta$ value with larger $N_S$.

The ground state expectation value of some operator $\mathcal{O}$ is calculated in the usual way from the Euclidean time functional integral

$$< \mathcal{O} > = \lim_{T \to \infty} \frac{1}{Z} \int \mathcal{D}\bar{q}(x) \mathcal{D}q(x) \mathcal{D}U_\mu(x) \; O(\bar{q},q,U) \; e^{-(S_U + S_F^W)} \tag{2.7}$$

which, after intergration of the quarks, yields

$$< \mathcal{O} > = \lim_{T \to \infty} \frac{1}{Z} \int \mathcal{D}U_\mu(x) \; det[M(U)] \; O(M^{-1}(U)) \; e^{-S_U} \quad . \tag{2.8}$$

We work in the *quenched* approximation where internal fermions loops are neglected, i.e. we set $det[M] = 1$ in (2.8). In this approximation the lattice QCD action $S_U$ couples only next neighbours and is thus amenable to Monte-Carlo methods. How good quenching is can only be assessed by doing an unquenched calculation and therefore one must bear in mind that all quenched results involve an unknown systematic error. At present unquenched calculations are starting to yield physical results and one can begin to examine the errors introduced by quenching. However the best evidence that the quenched approximation is reasonable comes from comparing recent quenched results [23] to experimental values. One finds that low lying hadron masses are within $6\% \pm 8\%$ and decay constants within $12\%$ to $20\%$ of experiment.

## 3. Static approximation

Going from light-light to heavy-light systems the problem we encounter has to do with the finite lattice spacing. We would like to study the B-meson and therefore we have to put on the lattice a u-quark and a b-(anti)quark with mass of about 4.5 GeV. Since we must keep linear $a$-terms small we require any mass to satisfy

$$a M < 1 \tag{3.9}$$

as well as

$$1/M < L \quad . \tag{3.10}$$

The second condition is imposed in order that the *finite volume* effects are small. For the B-meson we thus need

$$\frac{1}{a} > m_b \tag{3.11}$$
$$1/m_u < L$$

which is not yet within present computer capabilities.

Eichten [18] proposed a way to get around this problem. If one looks at the reduced mass, $\mu$, of the B-meson one realizes that it is about equal to the light quark mass $m_l$. Thus $\mu \lesssim \Lambda_{QCD}$ which is of course the reason why one can *not* apply a non-relativistic treatment. However this means that the b-quark mass or in general the heavy quark mass, $m_h$, is much larger that typical momenta for the heavy-light bound state. Therefore one can expand in powers of $1/m_h$. Keeping just the zeroth order term in this expansion is referred to as the static approximation. In this infinite mass limit the Dirac equation in an external field involves only the the zeroth component

$$(\gamma_0 D_0 - m_h) S_0(x,y) = \delta^4(x-y) \tag{3.12}$$

and solving for the static propagator $S_0(x,y)$ one obtains

$$S_0(x,y) = -i\delta^3(\vec{x}-\vec{y}) \; P\begin{pmatrix}x^0\\y^0\end{pmatrix} \left[ \frac{1+\gamma_0}{2} \; \Theta(x^0-y^0) \; e^{-im_h(x^0-y^0)} \right. \tag{3.13}$$
$$\left. + \; \frac{1-\gamma_0}{2} \; x^0 \longleftrightarrow y^0 \right] .$$

The static (stationary in space) propagator, as expected, has an upper (lower) component which propagates forward (backward) in time with an eikonal phase given by the path ordered exponential

$$P\begin{pmatrix}x^0\\y^0\end{pmatrix} = P\exp\left[ig_0 \int_{y^0}^{x^0} dz^0 A_0^c(\vec{x},z^0) t^c\right] \tag{3.14}$$

This static theory can now be put on the lattice. The heavy quark lattice propagator translates into

$$S_h(x;y) = \delta^3(\vec{x}-\vec{y}) \; \left\{ \Theta(x^0-y^0)\frac{1+\gamma_0}{2} U_0^\dagger(\vec{x};x^0-a) \cdots U_0^\dagger(\vec{x};y^0) \right. \tag{3.15}$$
$$\left. + \; \Theta(y^0-x^0)\frac{1-\gamma_0}{2} U_0(\vec{x};x^0) \cdots U_0(\vec{x};y^0-a) \right\},$$

In eq.(3.15), the exponential prefactor $\exp(-|x^0-y^0| m_h)$ has been dropped, since this corresponds only to a common shift of all energy levels by the bare mass of the heavy quark. In this way the mass of the heavy quark no longer appears and the condition $am_h < 1$ is relaxed.

One can envisage going one step further to take into account the first order $(1/m_h)$ terms. This leads to the familiar non-relativistic Lagragian

$$\mathcal{L}_{NRQCD} = Q^\dagger D_0 Q - Q^\dagger \left[\frac{\mathbf{D}^2}{2m_h} - c\,\frac{g_0}{2m_h}\sigma \cdot \mathbf{B}\right] Q \tag{3.16}$$

where $Q(x)$ is a two-component heavy quark; $c$ is one at tree level but gets renormalized, and therefore constitutes an additional parameter to be fixed. We remark here that NRQCD is non-renormalizable and a finite cut off has to be kept. A number of groups are investigating the applicability of (3.16) in the study of heavy-light and heavy-heavy systems [24, 25].

Our main approach is to use the static approximation as the leading term in the heavy quark expansion **and** to study propagating heavy quarks with masses less than $a^{-1}$ i.e. lighter than $m_b$. Interpolating between these results after taking their continuum limit we obtain the real estimate for $f_B$ [9, 16] or for other quantities of the b-system like the mass of $\Lambda_b$ [19].

## 4. Pseudoscalar Decay Constant

The pseudoscalar decay constant $f_P$ is extracted from the lattice matrix element through the relation

$$< 0|\mathcal{M}^{loc}_{\gamma_0\gamma_5}|P> = \frac{1}{Z_A}\sqrt{M_P/2}\, f_P\, a^{3/2} \quad . \tag{4.17}$$

$Z_A$ is the axial current renormalization constant, $\mathcal{M}^{loc}_{\gamma_0\gamma_5}$ is the zeroth component of the local axial vector current, and $|P>$ denotes the pseudoscalar ground state. The renormalization factor $Z_A$ will be different for propagating heavy quarks and for the static theory. The latter we denote by $Z_A^{stat}$ and required a separate perturbative calculation [26]. We define a lattice current by

$$\mathcal{M}^J_{\gamma_0\gamma_5}(\vec{x},t) = \bar{h}(\vec{x},t)\, \gamma_0\gamma_5\, l^J(\vec{x},t) \tag{4.18}$$

where

$$l^J(\vec{x},t) = \sum_{\vec{y}} \Phi^J(\vec{x},\vec{y},\mathcal{U}(t))\, l(\vec{y},t) \tag{4.19}$$

is a smeared light ($l$) quark field obtained by convoluting with the trial wave function $\Phi^J$ and $h(\vec{x},t)$ is the local heavy ($h$) quark field.

The aim is to extract the local matrix element by using wave functions optimized to yield early ground state dominance in the smeared meson-meson correlator

$$C^{I,J}_{\gamma_0\gamma_5}(t) = \sum_{\vec{x}} <\mathcal{M}^I_{\gamma_0\gamma_5}(\vec{x},t)\, [\mathcal{M}^J_{\gamma_0\gamma_5}(\vec{0},0)]^\dagger> \quad . \tag{4.20}$$

A way to extract the *local* matrix element $<0|\mathcal{M}^{loc}_{\gamma_0\gamma_5}|P>$ from the local-smeared, $C^{loc,J}_{\gamma_0\gamma_5}$, and smeared-smeared, $C^{I,J}_{\gamma_0\gamma_5}$, correlator, is to do a fit to the ratio

$$R(t) = \frac{C^{loc,J}_{\gamma_0\gamma_5}(t)}{\sqrt{C^{J,J}_{\gamma_0\gamma_5}(t)}} \stackrel{t\,\text{large}}{\rightarrow} <0|\mathcal{M}^{loc}_{\gamma_0\gamma_5}|P>\, e^{-M_P t/2} \quad . \tag{4.21}$$

Ground state dominance is monitored by the plateau in the local mass

$$\mu_R(t) = \ln\frac{R(t)}{R(t-a)} \quad . \tag{4.22}$$

In the static approximation ground state dominance has proven particularly difficult. Unsuspected contamination from excited states was partly responsible for the large values of $f_B^{\text{stat}}$ obtained at the initial stages. To cure this problem various wave functions have been used by different groups [15, 27]. We chose to construct gauge covariant wave functions so that no problems due to gauge fixing occur. After experimenting with the parameters and form of these wave functions we found that an optimum choice to obtain early plateaus in both smeared-smeared and local-smeared correlators occurs for wave functions with r.m.s. radius of approximately 0.3 fm. This is a quite reasonable size for a hadronic wave function. In the following we will use the 'best' wave function, i.e. a 'gaussian type' wave function with parameters $n = 100$ and $\alpha = 4$. We studied lattices at $\beta = 5.74, 6.00$ and $6.26$. At each $\beta$ value we varied the spatial size to investigate the volume dependence.

## 4.1 Results in the static approximation

We consider the quantity

$$\hat{F} = f_P \sqrt{M_P} \left( \frac{\alpha_s(M_P)}{\alpha_s(M_B)} \right)^{2/11} . \tag{4.23}$$

which has a well defined limit as $M_P \to \infty$ [8]. In fig. 2 we display the dependence of $\hat{F}$ scaled by the string tension $\sigma$ as a function of the spatial length of the lattice also in units of $\sigma$.

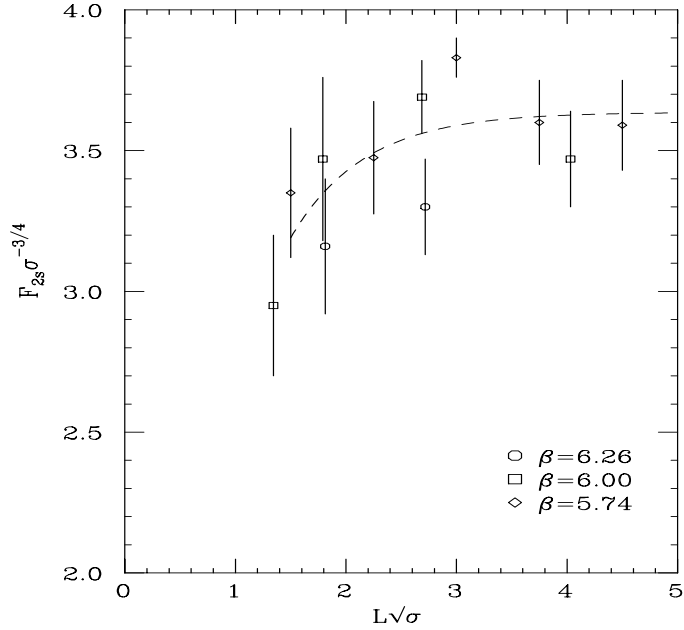

Figure 2: $\hat{F}\sigma^{-3/4}/Z_A^{stat}$ vs $L\sigma^{1/2}$ at $\beta = 5.74, 6.0$ and $6.26$. The light quark mass is fixed at about twice the strange quark mass. The dashed line is the result of fitting all data to the form $C_0 - C\exp(-1.5L\sigma^{1/2})$.

From this figure we conclude that for $L \gtrsim 1$ fm ($L\sqrt{\sigma} \gtrsim 3$) the volume effects are small in comparison with our statistical errors. Fixing our volume at $L \sim 1$ fm we then study finite-$a$ effects which we expect to be the largest source of systematic error.

At $\beta = 6.0$ and using the lattice spacing $a = (2.2 \text{ GeV})^{-1}$ obtained from Allton's analysis [14] we display in fig. 3 the dependence of $\hat{F}$ as we increase the quark mass of the light propagating quark from about the strange quark mass to about half the charm mass.

To obtain the decay constant for the physical mesons we need to extrapolate the light quark mass to the chiral limit. This is done in the standard way by first determining the critical $\kappa$ value, $\kappa_c$, as explained in section 2. $\hat{F}$ is extrapolated to $\kappa_c$ by fitting linearly to $M_P^2(l,l)$ for the three lightest quark masses as shown in fig. 3. The results at the chiral limit for our three $\beta$ values are shown in fig. 4.

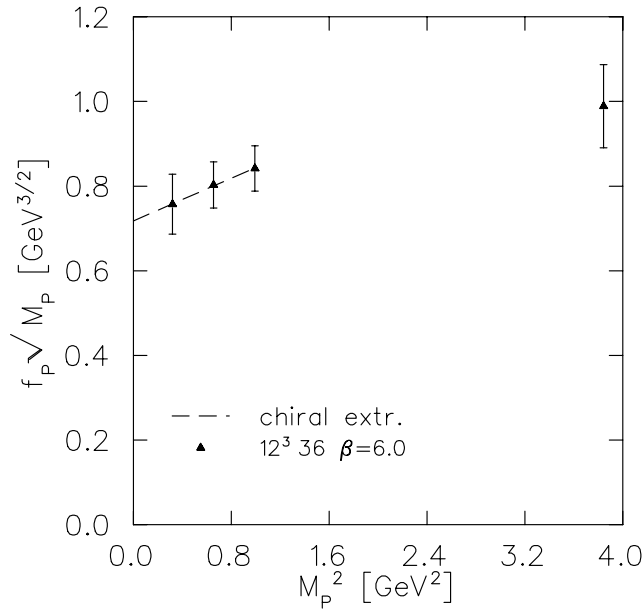

Figure 3: $\hat{F}$ in GeV $^{3/2}$ in the static approximation is shown vs the pseudoscalar light-light quark mass, $M_P^2(l,l)$, in GeV$^2$ at $\beta = 6.0$, where we took $a^{-1} = 2.2$ GeV from ref.[14].

Following ref. [14] we use for the continuum extrapolation the results at the two largest $\beta$ values. This yields $\hat{F} = 0.64(4)$ GeV$^{3/2}$. If we use the continuum extrapolation of $f_\pi/\sqrt{\sigma} = 0.270(12)$ and of $\hat{F}/\sigma^{3/4} = 1.95(22)(22)$ from ref. [9, 16] we obtain $\hat{F} = 0.67(7)(7)$ GeV$^{3/2}$ consistent with that obtained using the $a$ values from Allton's analysis. On the other hand, if one uses the string tension to set the scale the value goes down to $0.53(6)(6)$ GeV$^{3/2}$ which however is still within errors. The value at the B-meson mass translates to

$$205 \lesssim f_B^{\mathrm{stat}} \lesssim 296 \quad \mathrm{MeV} \qquad (4.24)$$

corresponding to the lower and upper values obtained from the string tension and from Allton's values of $a$ respectively (only the statistical errors were included in this bound). The Fermilab group has recently reported a value of $188(23)(15)(+26)(14)$ MeV [13], where with their good wave functions the isolation of the ground state was optimal. In fig. 5 we summarise the results of several groups. Apart from some high values at $\beta = 6.0$ coming from older calculations, there is an overall agreement. In the same figure we display recent results from sum rules [28] also in the static limit which show a range of values in agreement with the lattice results.

The dependence of $f_B^{stat}$ on the light quark mass is best obtained from ratios where systematic errors due to scale and renormalization cancel. Increasing the mass to the strange quark mass we obtain in the continuum limit,

$$\frac{f_{B_s}^{stat}}{f_B^{stat}} = 1.14(11) \quad . \qquad (4.25)$$

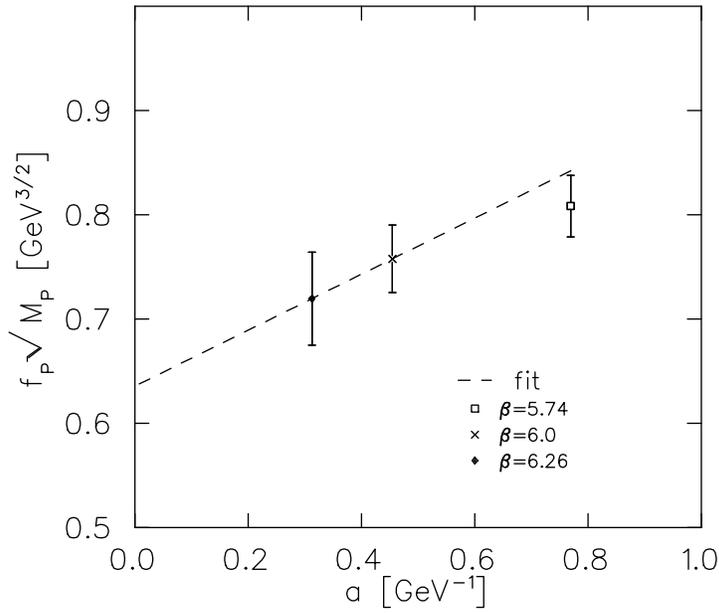

Figure 4: $\hat{F}$ in GeV$^{3/2}$ in the static approximation at $\beta = 5.74, 6.0, 6.26$ vs $a$ in GeV$^{-1}$ taken from ref. [14]. The dashed line indicates what the continuum limit is.

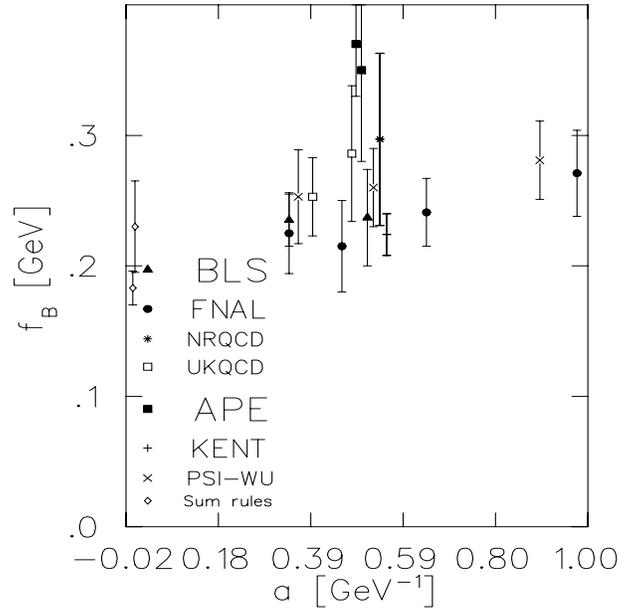

Figure 5: $f_B$ in GeV in the static approximation as a function of the lattice spacing $a$ in GeV $^{-1}$ from various groups ( BLS[11], FNAL[13], NRQCD[24], UKQCD[12], APE[10], KENT[29], PSI-WUP[9], Sum rules[28]).

## 4.2 Results using propagating heavy quarks

How good are the results obtained in the static limit? We know that at D-meson mass the $1/m_h$ corrections are as large as 40%. In order to answer this question for the B-meson we have to do the calculation with propagating heavy quarks on the lower side of the b-quark mass and try to interpolate to the static result. For the lattices that we used we can reach masses of about 2.5 GeV.

Finite volume effects are on similar ground as with the static theory and will contribute a 4% systematic error. To filter out the ground state we will again smear the light quark as we have done in the static approximation. Our most serious concern here is the finite lattice resolution. When we are close to the continuum limit dimensionless ratios should scale, i.e. should show no dependence on the scale. When the dependence on $a$ is weak one may extrapolate linearly as was done for the static results. In the $f_B$ case however the $a$-dependence is stronger and such an extrapolation is no longer justified. However, what one can still do, is to extrapolate the results for lighter masses to the continuum since the $a$-dependence is weaker there. Therefore we can obtain continuum results for pseudoscalar masses in the range $1 - 2.3$ GeV. These are shown in fig. 6 together with the static continuum result.

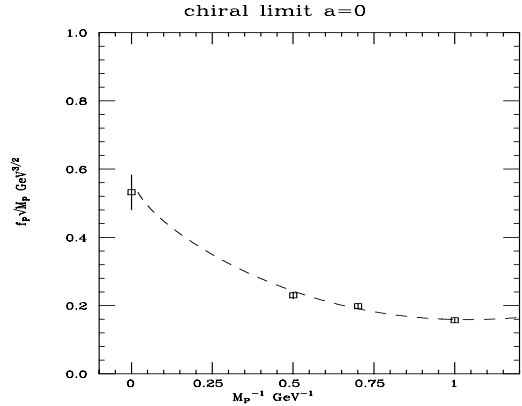

Figure 6: $\hat{F}$ in GeV$^{3/2}$ vs $1/M_P$ in GeV$^{-1}$ at the chiral limit and in the continuum.

We fit the results at finite values of $M_P$ and the static result to a power series in $1/M_P$. Using $c_0 + c_1/M_P + c_2/M_P^2$ we obtain at the B-meson mass

$$f_B = 180(50) MeV \qquad (4.26)$$

This result is in agreement with results from other groups, as shown in fig. 7, where we also display results coming from sum rules [28, 30].

The dependence on the light quark mass is again best given in the form of the ratio

$$f_{B_s}/f_B = 1.14(5) \qquad (4.27)$$

Fig. 8 shows the experimentally measured decay constants for the light mesons and the lattice results filling the gap in the heavy-light region.

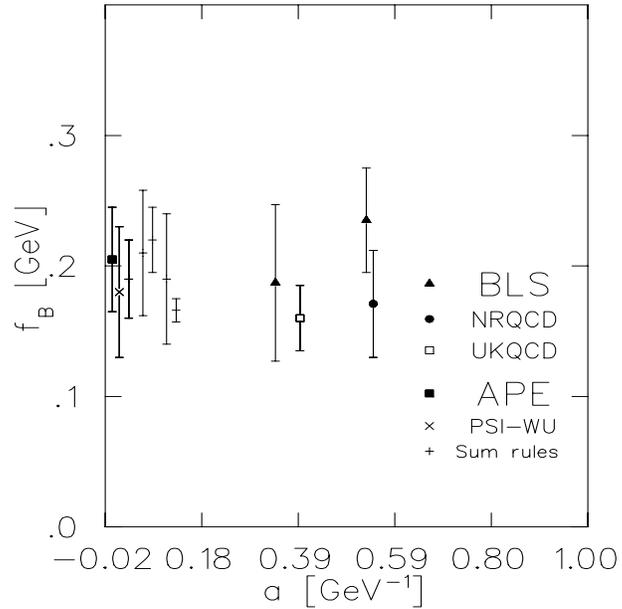

Figure 7: $f_B$ in GeV with propagating heavy quarks vs $a$ in GeV$^{-1}$. The sum rule results were taken from ref. [6,28,30] and are horizontally displayed from $a = 0$ for clarity (BLS[11], NRQCD[24], UKQCD[12], APE[15], PSI-WU[16]).

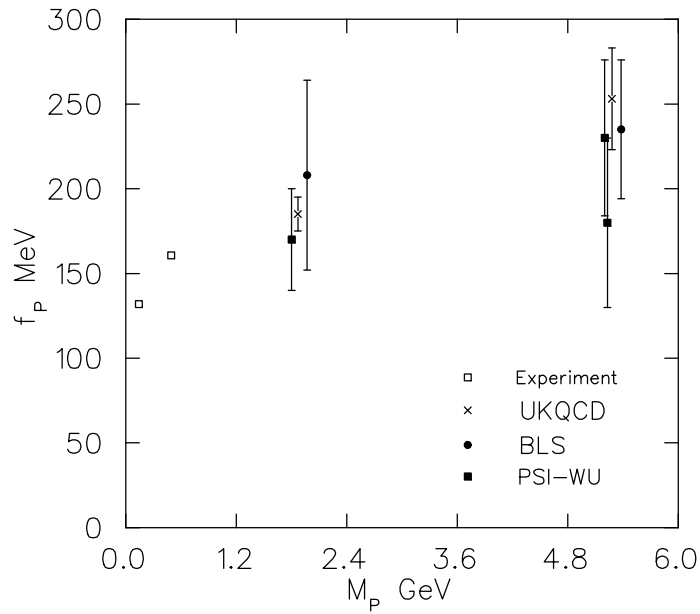

Figure 8: The pseudoscalar decay constant $f_P$ in MeV vs the meson mass in GeV. At the B-meson mass the data referred to as UKQCD[12], BLS[11] and the higher value from PSI-WU[9] are static results, whereas the lowest value from PSI-WU[16] is obtained using propagating heavy quarks.

# 5. Beautiful Baryons

We use the same idea of interpolating results from lower mass heavy propagating quarks, that we used for $f_B$, to study the baryonic sector. We calculate the mass of $\Lambda_b$, a baryon made of one b-quark and two light quarks. We avoid computing the mass of the $\Lambda_b$ directly, but rather calculate the mass splittings $\Delta_\Lambda = M_{\Lambda_b} - M_B$ with respect to the B-meson mass. These splittings do not depend on the heavy quark mass in the infinite mass limit and are therefore less prone to contamination by finite $a$ effects in the $b$ and $c$ quark mass regions. The interpolating field for the $\Lambda$ baryon [31] is taken as

$$C_\Lambda(t) = \sum_{\vec{x}} \left\langle \left(\epsilon^{abc} h_a(x) \left(u_b^J(x) C\gamma_5 d_c^J(x)\right)\right) \left(\epsilon^{abc} h_a(0) \left(u_b^J(0) C\gamma_5 d_c^J(0)\right)\right)^\dagger \right\rangle, \qquad (5.28)$$

where smearing is applied to the light quarks $u$ and $d$.

Given the lattice results for the $\Lambda$ correlator, $C_\Lambda$, and the pseudoscalar correlator, $C_P = \sum_{\vec{x}} < (\bar{h}(x)\gamma_5 l^J(x))(\bar{l}(0)\gamma_5 h(0)) >$, we perform a direct fit to their ratio

$$R_\Lambda(t) = \frac{C_\Lambda(t)}{C_P(t)} \to A e^{-\Delta_\Lambda t} \qquad (5.29)$$

which in the large $t$ limit yields the mass splitting. In fig. 9 we show our results at the three $\beta$ values together with the result from UKQCD which used the clover action. The $a$-dependence is well within the statistical error and we therefore make a $1/M_P$ fit extrapolating the low mass data to the B-meson mass. On the same figure we display the static results at $1/M_P = 0$. The highest point is the result of an earlier calculation [31] whereas the lowest is a recent UKQCD result using the clover action [32]. The middle point is our result [9].

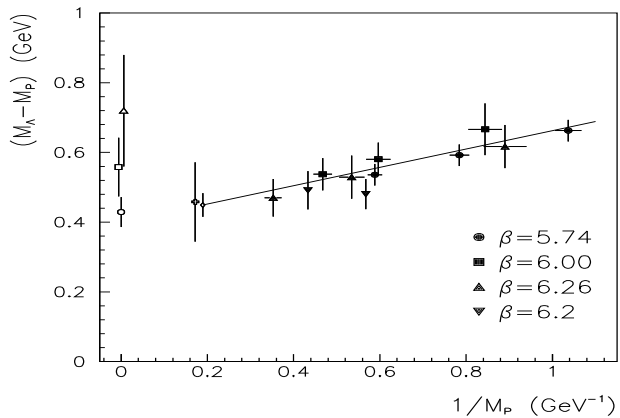

Figure 9: The $\Lambda_b - M_P$ mass splitting in GeV vs $1/M_P$ in GeV$^{-1}$. The data at $\beta = 6.2$ come from UKQCD[33]. The results at $1/M_P = 0$ are static results from [31],[9] and [32] in order of decreasing size.

Using the experimental value $M_B = 5.27$ GeV we obtain

$$M_{\Lambda_b} = 5.728 \pm 0.144 \pm 0.018 \text{ GeV} \qquad (5.30)$$

in good agreement with the result coming from the UA1 collaboration.

## 6. Phenomenological consequences

We briefly comment about possible phenomenological consequences on the expected size [5, 33] of CP violation effects in B-decays, using our value of $f_B = 180(50)$ MeV.

As usual, we start from the unitarity triangle

$$\sum_{i=u,c,t} V_{id} V_{ib}^* = 0$$

associated with the $b \to d$ transitions. Using the standard Wolfenstein parametrization we have $V_{ub}^* - |V_{cd}V_{cb}^*| + V_{td} = 0$ to a very good approximation. After rescaling by $|V_{cd}V_{cb}^*|^{-1}$ this defines a triangle in the complex plane with vertices $(\rho, \eta)$, $(1,0)$ and $(0,0)$ where

$$\rho = (\text{Re } V_{ub})/|V_{cd}V_{cb}^*|, \quad \eta = -(\text{Im } V_{ub})/|V_{cd}V_{cb}^*| \ . \tag{6.31}$$

We use the updated experimental constraints as given in [34]:

$$|V_{ub}/V_{cb}| = |V_{us}|\sqrt{\rho^2 + \eta^2} = 0.08 \pm 0.03 \tag{6.32}$$

$$\text{with } |V_{cb}| = 0.039 \pm 0.006 \text{ and } \sqrt{\rho^2 + \eta^2} = 0.36 \pm 0.14 \ ,$$

the $K^0 - \bar{K}^0$ CP-violating parameter,

$$|\epsilon| = 4.00 \times 10^4 \ B_K |V_{cb}|^2 |V_{us}|^2 \eta \left[ F(x_c, x_t) + |V_{cb}|^2 (1 - \rho) F(x_t) \right] = (2.26 \pm 0.02) \times 10^{-3} \tag{6.33}$$

and the $B^0 - \bar{B}^0$ mixing parameter

$$x_d = 4.15 \times 10^{-7} \ B_B \tau_B m_B^{-1} f_B^2 |V_{cb}|^2 |V_{us}|^2 \left((1 - \rho)^2 + \eta^2\right) F(x_t) = 0.716 \pm 0.04 \ . \tag{6.34}$$

We take $\tau_B = 1.63(7) \ ps$ and following ref.[34] we assume $B_K = 0.8 \pm 0.2$ and take $B_B = 1$. The functions $F$ are the QCD corrected versions of the Inami-Lim functions depending on the charm mass $x_c = m_c^2/M_W^2$ and the top mass $x_t = m_t^2/M_W^2$ [35]. It is obvious from the expression for $x_d$ that a large (small) value for $f_B$ favours positive (negative) $\rho$.

Fig. 10 shows the allowed $(\rho, \eta)$ range for a top mass $m_t = 174$ GeV, using the above phenomenological constraints.

From this figure it is clear that with the present constraints we are not able to pin down the unitarity triangle. However there is a good possibility for a positive value of $\rho$ which will be good news for CP asymmetry.

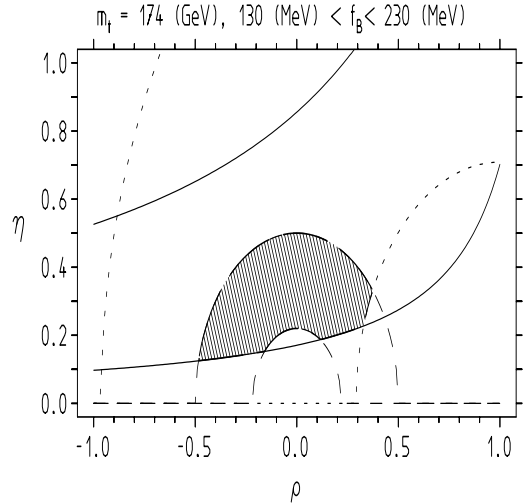

Figure 10: The shaded region is the allowed region for the unitarity triangle.

## 7. Conclusions

Within the quenched approximation we have been able to perform a detailed study of the limiting behaviour of $f_P\sqrt{M_P}$ by varying the heavy quark mass $m_h$ and the lattice spacings $a$ in suitable ranges.

By means of extrapolation, we were able to obtain the decay constant of the B-meson, $f_B = 180(50)$ MeV, as well as the mass of the $\Lambda_b$, $M_{\Lambda_b} = 5.728(144)(18)$ GeV. Finite lattice spacing effects still constitute the largest source of systematic error accounting for most of our errors in the lattice results.

In view of the phenomenological importance of $f_B$ to CP violation in the CKM scenario, it would be highly desirable to tighten the bounds set in eq. (6.34). This would require pushing $\beta$ to higher values but in addition using an $a$-improved action.

**Acknowledgements.** I enjoyed the hospitality of the Paul Scherrer Institute, where part of this talk was prepared. This work was done in an enjoyable collaboration with A. Borrelli, F. Jegerlehner, S. Güsken, K. Schilling and R. Sommer.

# References

[1] D. G. Hitlin, in *Proceedings of the Snowmass Summer Study on High Energy Physics*, Snowmass, Colorado 1990.

[2] N. Cabibbo, Phys. Lett. (1963) 513; M. Kobayashi and K. Maskawa, Prog. Theor. Phys. 49 (1973) 652.

[3] P. Franzini, Phys. Rep. C 173 (1989) 1; E. A. Paschos and U. Türke, Phys. Rep. C 178 (1989) 145; T. Nakada, in *International Symposium on Lepton and Photon Interactions*, Ithaca, NY, 1993, PSI report, PSI-PR-93-18.

[4] H. Albrecht et al (ARGUS Collaboration), Phys. lett. B 324 (1994) 249; B. Barish et al (CLEO Collaboration), Preprint CLNS94/1285 (1994); D. Buskulic et al (ALEPH Collaboration), in *International Conference on High Energy Physics*, Glasgow, ICHE94-0605 (1994).

[5] C. Dib, I. Dunietz, F. Gilman and Y. Nir, Phys. Rev. D41 (1990) 1522.

[6] T. M. Aliev and V. L. Eletsky, Sov. J. Nucl. Phys.38 (1983) 936; L. J. Reinders, H. Rubinstein and S. Yazaki Phys. Rep. C127 (1985) 1; C. A. Dominguez and N. Paver, Phys Lett. B197 (1987) 423; B246 (1990) 493; S. Narison, Phys. Lett. B98 (1987) 104; Phys. Lett. B218 (1989) 238; A. Pich, Phys. Lett. B206 (1988) 322.

[7] M. Suzuki, Phys. Lett. B162 (1985) 392; D. Silverman and H. Yao, Phys. Rev. D38 (1988) 214; S. Capstick and S. Godfrey, Phys. Rev. D41 (1990) 2856.

[8] E. V. Shuryak, Nucl. Phys. B198 (1983) 83; M. A. Shifman and M. B. Voloshin, Sov. J. Nucl. Phys. 45 (1988) 292; H. D. Politzer and M. B. Wise, Phys. Lett. B208 (1988) 504.

[9] C. Alexandrou, S. Güsken, F. Jegerlehner, K. Schilling and R. Sommer, Phys. Lett. B256 (1991) 60; Nucl. Phys. B414 (1994) 815; R. Sommer, preprint DESY-94-011.

[10] The APE collaboration, C. R. Allton et al, preprint CERN-TH-7202-94; Nucl. Phys. B413 (1994) 461.

[11] C. W. Bernard, Nucl. Phys. B (Proc. Suppl.) 34 (1994) 47; C. W. Bernard, J. N. Labrenz and A. Soni, Phys. Rev. D49 (1994) 2536.

[12] UKQCD Collaboration (R. M. Baxter et al), Phys. Rev. D49 (1994) 1594.

[13] A. Duncan, E. Eichten, J. Flynn, B. Hill, G. Hochney, H. Thacker, preprint FERMILAB-PUB-94-164-T; E. Eichten, Ch. Quigg, Phys. Rev. D49 (1994) 5845.

[14] C. R. Allton, preprint ROME-1041-94.

[15] The European Collaboration, (A. Abada et al), Nucl. Phys. B 376 (1992) 172.

[16] C. Alexandrou, S. Güsken, F. Jegerlehner, K. Schilling and R. Sommer, Z. Phys. C62 (1994) 659.

[17] UKQCD Collaboration (C. R. Allton et al), Phys. Rev.D 49 (1993) 474; Nucl. Phys. B407 (1993) 331; APE Collaboration, (C. R. Allton et al), Phys. Lett. B326 (1994) 295.

[18] E. Eichten, in *Field Theory on the Lattice*, Nucl. Phys. B (Proc. Suppl.) 4 (1988) 147.

[19] C. Alexandrou, A. Borrelli, S. Güsken, F. Jegerlehner, K. Schilling, G. Siegert and R. Sommer, Phys. Lett B337 (1994) 340; Lattice '94 Bielefeld, Germany, to appear in Nucl. Phys. B Proc. Suppl.

[20] K. G. Wilson, Phys. Rev. D10 (1974) 2445; *New Phenomena in Subnuclear Physics*, Erice 1975, Plenum, New York (1977).

[21] M. Creutz in *International Conference on Quark Confinement and the Hadron Spectrum*, Como, Italy, 1994; A. A. Slavnov, these proceedings.

[22] J. Gasser and H. Leutwyler, Phys. Rep. C87 (1982) 77.

[23] F. Butler, H.Chen, J. Sexton, A. Vaccarino and D. Weigarten, Phys. Rev. Lett. 70 (1993); Nucl. Phys. B421 (1994) 217; IBM preprint IBM-HET-94-3.

[24] S. Hashimoto, Phys. Rev. D50 (1994) 4639; Nucl. Phys. B (Proc. suppl.) 34 (1994) 441.

[25] UKQCD Collaboration, (C. T. H. Davies *et al*), Nucl. Phys. B (Proc. Suppl.) 34 (1994) 437.

[26] Ph. Boucaud, C. L. Lin, and O. Pene, Phys. Rev. D40 (1989) 1529;E.Eichten, B. Hill, Phys.Lett.B240(1990)193; Ph. Boucaud, J. P. Leroy, J. Micheli, O. Pene, and G. C. Rossi, Phys. Rev. D47 (1993) 1206

[27] S. Güsken, Nucl. Phys. B17 (Proc. Suppl.) (1990) 361

[28] E. Bagan, P. Ball, V. M. Braun, H. G. Dosch, Phys. Lett. B278 (1992) 464; V. Eletsky and E. Shuryak, Phys. Lett. B276 (1992) 191.

[29] T. Draper and C. McNeile, Nucl. Phys B (Proc. Supp.) 34(1994) 453.

[30] M. Neubert, Phys. Rev. D45 (1992) 2451.

[31] M. Bochichio, M. Martinelli, C. R. Allton, C.T. Sachrajda and D. B. Carpenter, Nucl. Phys. B372 (1992) 403.

[32] UKQCD Collaboration, in Lattice '94, Bielefeld, Germany.

[33] M. Lusignoli, L. Maiani, G. Martinelli and L. Reina, Nucl. Phys. B369 (1992) 139.

[34] Ali and D. London, preprint CERN-TH.7398-94.

[35] T. Inami and C. S. Lim, Prog. Theor. Phys. 65 (1981) 297; A. J. Buras, M. Janin and P. H. Weisz, Nucl. Phys. B347 (1990) 491.